# Revisiting the crystal structure of the equilibrium S (Al$_2$CuMg) phase in Al-Cu-Mg alloys using X-ray absorption spectroscopy (XAFS)

*Danny Petschke[a], Frank Lotter[b], Torsten E. M. Staab[c]

a) University Wuerzburg, Department of Chemistry and Pharmacy,
   LCTM Roentgenring 11, D-97070 Wuerzburg, Germany
   **email:** danny.petschke@uni-wuerzburg.de) – *corresponding author

b) University Wuerzburg, Department of Chemistry and Pharmacy,
   LCTM Roentgenring 11, D-97070 Wuerzburg, Germany
   **email:** frank.lotter@uni-wuerzburg.de

c) University Wuerzburg, Department of Chemistry and Pharmacy,
   LCTM Roentgenring 11, D-97070 Wuerzburg, Germany
   **email:** torsten.staab@uni-wuerzburg.de

## Abstract

Even though, the crystal structure of the intermediate (S') and the equilibrium S (Al$_2$CuMg) phase were subject of many investigations by using mostly imaging or diffraction techniques, the results remain still controversial. In this study, we used X-ray absorption spectroscopy (XAFS) to verify the correct crystal structure considering the well-known models reported by Perlitz & Westgren (PW), Mondolfo, Radmilovic & Kilaas and Yan *et al.* The S phase structure was confirmed by direct comparison to simulated XAFS spectra using FDMNES. Our results support the widely accepted PW model as the correct structure while other models do not match our observations.

## 1      Introduction

Al-Cu-Mg alloys are the basis of the AA2xxx series of age-hardenable aluminium alloys, which find wide applications in the aviation sector because of their low specific weight combined with their high mechanical strength and susceptibility to corrosion.

The classic precipitation sequence of Al-Cu-Mg alloys was first proposed in 1952 by Bagaryatsky [1,2]:

SSSS → GPB zones → S'' (GPB-II zones) → S' → equilibrium S (Al$_2$CuMg)

where *SSSS* is the supersaturated solid solution and *GPB* stands for *Guinier-Preston-Bagaryatsky* zones (termed by Silcock [3]), which Bagaryatsky suggested as short-range ordered Cu and Mg atoms populating the {100}$_\alpha$ planes. Whilst this sequence has been cited by many authors, the crystal structure of the phases resulted in controversial discussions and the existence of metastable S phase variants (GPB-II/S'', S') remains a relevant topic in research [4–12].

For the purposes of this publication, we do not distinguish between metastable S' phase and equilibrium ($Al_2CuMg$) S phase precipitates, since many authors reported that they mainly differ in their level of coherence and orientation relationship with the Al-matrix while otherwise their structures seem to be identical [5,6,9–11,13]. Thus, all proposed structures [4,14–17] as well as five recently published refinements [10–12], which clearly refer to the intermediate (S') or equilibrium S phase were considered in this study and, hence, are described uniformly as *S phase* hereafter. Structures, which are proposed as S'' (GPB-II) phase such as by Cuisat *et al.* [18], Wang & Starink [19], Kovarik *et al.* [20–22] or Wolverton [23] are not taken into consideration here even though some authors state that their structure is similar to the (equilibrium) S phase [24] or that they even coexist with S' and S phase precipitates [25].

As listed in Table 1 and 2, several models have been proposed for the crystal structure of the S phase. First in 1943, by using X-ray diffraction (XRD) on small single crystals grown from the melt, Perlitz & Westgren (*PW*) reported a centered orthorhombic crystal structure (space group: Cmcm) with 16 atoms per unit cell and with lattice parameters a = 4.0Å, b = 9.23Å and c = 7.14Å [14]. Nowadays, this model is the widely accepted in the Al community but, nonetheless, measurements on the crystal structure and morphology are still controversially discussed:

In 1976, Mondolfo proposed a modified version of the *PW* model (space group: P1) with slightly different lattice parameters and one $\{001\}_S$ Cu-Mg layer shifted along the $[010]_S$ direction [15]. However, first principle calculations by Liu *et al.* showed that the Mondolfo model is energetically unstable without transforming to the *PW* model [26]. In 1990, based on HRTEM measurements, Yan *et al.* proposed an orthorhombic structure with 4 atoms per unit cell (space group: Pmm2) and lattice parameters a = 4.0Å, b = 4.61Å and c = 7.18Å [16]. Moreover, they found no match to the Mondolfo model by comparing simulations to the dynamic diffraction pattern. X-ray powder diffraction measurements on two different Al-Li-Cu-Mg alloys by Pérez-Landazábal *et al.* interestingly supported the model of Mondolfo and rejected the structure proposed by Yan *et al.* [27]. In 1999, Radmilovic *et al.* critically evaluated the models of Mondolfo, Yan *et al.* and Perlitz & Westgren and found that only the *PW* model gave consistent results to their HRTEM images [4]. In addition, they suggested a new structure, often called *RaVel* model (hereafter called *RKDS*), which is identical to the *PW* model but with Cu and Mg interchanged [4]. In 2001, Kilaas & Radmilovic refined the atomic positions of their latter model [17]. Nevertheless, this model was first rejected by Wolverton in 2001, who determined a high positive formation enthalpy (+16.4 kJ $mol^{-1}atom^{-1}$) as a result of first principle calculations [23]. A positive formation energy indicates that this structure is energetically unstable. However, the calculated formation enthalpy of the *PW* model (-19.4 kJ mol$^{-1}$atom$^{-1}$) is in a good agreement with the experimentally obtained value (-15.8 kJ mol$^{-1}$atom$^{-1}$ [28]). In 2004, Majimel *et al.* supported the structure proposed by Radmilovic & Kilaas since simulations of this model matched their HRTEM images better than the *PW* and Mondolfo model [7]. However, in 2011, Liu *et al.* obtained results for the *RKDS* model (+15.9 kJ mol$^{-1}$atom$^{-1}$) [26] by first principle calculations, which are fully consistent with those by Wolverton (+16.4 kJ mol$^{-1}$atom$^{-1}$) [23].

Even though, the *PW* model was favoured by several authors before using mostly diffraction and/or imaging techniques [4,7,9–12,15–17,26,27], the goal of this study is to verify the *true* S phase model in a pure ternary Al-Cu-Mg alloy by applying a different approach: X-ray absorption fine structure spectroscopy (XAFS). This rarely used but powerful method allows us to probe the local atomic surroundings of one specific absorbing element (here Cu) [29,30] and, hence, provides a high sensitivity to changes in the crystal symmetry and the atomic site occupancies of Cu and Mg atoms forming the S phase. The first application on binary Al-Cu alloys appeared in 1979 by Fontaine *et al.* [31]. In 2011, Staab *et al.* presented a reasonable agreement to the *PW* model by comparing calculated FEFF spectra to the experimentally obtained curve from an Al-Cu-Mg alloy aged for one hour at 350°C [32]. However, the authors did not undertake a comparison to other structures proposed for the S phase.

## 2 Methodology

### 2.1 Experimental procedure

#### 2.1.1 Materials and heat treatments

In this study, all samples have been prepared from the same as-cast and homogenized (4h at 515°C) high purity alloy having a nominal composition of Al - 2.2wt.% (1.0at.%) Cu - 1.6wt.% (1.8at.%) Mg. This alloy clearly lies in the two-phase field of the ternary phase diagram ((fcc)α + S) and, thus, the formation of S phase precipitates is assured [33]. The material is prepared into samples of appropriate size for the different methods, solution heat treated for 1h at 515°C under flowing argon gas, quenched into ice-water (called *AQ* hereafter) and artificially aged for one day at 200°C to assure that mainly S phase precipitates are formed. For DSC measurements, the square-shaped samples (~45mg) were polished at one side to assure a consistent contact surface to the Al-crucibles. The *AQ* sample was measured within 2-3min after quenching. For XAFS measurements, samples with a diameter of 10mm were thinned to approximately 100-200µm.

#### 2.1.2 Differential scanning calorimetry (DSC)

The DSC measurements were carried out in a *Netzsch 204 F1 Phoenix* apparatus (heat-flux type) under nitrogen atmosphere with a heating rate of 20Kmin$^{-1}$ from -20°C up to 530°C. The apparatus baseline was measured under the same conditions by using pure Al (5N) as both sample and reference. Subsequently, the raw data were normalized by mass and corrected for the baseline to provide a meaningful comparison between the two curves (Fig. 1).

### 2.1.3 X-ray absorption fine structure spectroscopy (XAFS)

The XAFS measurements were carried out on the ID24-L beamline at the European Synchrotron Radiation Facility (ESRF) in Grenoble, France, which is designed for time-resolved purposes on the millisecond scale. The spectra were recorded at the Cu-K edge (8979eV [34]) under the same conditions as described by the authors in a previous study [35].

### 2.2 Simulated structures

The atomic positions, which served as input for the spectra simulation, are listed in Table 1 and 2. For the spectra simulation we used *FDMNES* [36] in the *muffin-tin approximation* for the atoms (keyword: *GREEN*). Furthermore, the resulting spectrum of the linearly combined absorbing Cu atoms was convoluted by an energy dependant broadening described by an *arctangent function* [36] (keyword: *CONVOLUTION*). Moreover, the simulation radius around the scattering atoms was determined from simulated spectra of pure Cu with iteratively increasing radii. Therefore, the radius was chosen to be 900pm since no significant changes in the XANES region (~8980-9100eV) occur anymore. Finally, all simulated and experimentally obtained spectra were normalized using the *Demeter* software package *ATHENA* [37].

The structures denoted as *M_1* (*M_2*) [15], *YCM* [16] and *RKDS_1* (*RKDS_2*) [4,17] represent the proposed structures, which significantly deviate in symmetry and/or the atomic positions from the widely accepted *PW* model (Fig. 2a). On the other hand, *HHP* [12], *SHCDB_1* (*SHCDB_2*) [10] and *SHCDB_3* (*SHCDB_4*) [11] assign recent refinements of the *PW* model, which have the same symmetry group but slight modifications for lattice parameters and atomic positions (Fig. 2b).

The *RKDS_1* model refers to the first structure proposed by Radmilovic *et al.*, which is similar to the *PW* model (except for the interchange of Cu and Mg and modifications in the lattice parameters) [4], whereas the *RKDS_2* model represents the refined *RKDS_1* model with modified fractional atomic site positions [17].

The refined *PW* structures by Styles *et al.* denoted as *SHCDB_1 (SHCDB_3)* and *SHCDB_2 (SHCDB_4)* refer to the intermediate (S') phase and equilibrium S phase structure, respectively. The data were taken from two different studies of the years 2012 and 2015 [10,11].

### 3 Results and discussion

First, DSC measurements were carried out to verify the presence of S phase precipitates in the studied sample state (200°C, 1d):

As seen in the AQ scan (red curve, Fig. 1), two major exothermic peaks at ~75°C and 250-300°C can be detected and refer to the formation of GPB zones (and/or Cu-Mg co-clusters[1]) and S phase precipitates, respectively [25,38–40]. After aging at 200°C for one day (blue curve, Fig. 1), a flat curve in the temperature region of 25-250°C (grey shaded box, Fig. 1) appears and evidences the absence of GPB (co-clusters) and GPB-II zones, which typically dissolve at temperatures between 200°C and 250°C [38,40]. Furthermore, the S formation peak (250-300°C) has completely disappeared, which indicates the transformation of the majority of metastable phases into S phase precipitates. This observation is consistent with the studies of Wang & Starink who confirmed the completion of S phase formation after 12h at 190°C using TEM with SAD on the AA2024 alloy [38].

At the first glance, by comparing the experimentally obtained XAFS spectrum to the simulated spectra of the proposed structures in Fig. 2a, one can clearly identify a good agreement with the widely accepted *PW* and *M_1* (*M_2*) models. However, the simulated spectra of the structures proposed by Radmilovic & Kilaas (*RKDS_1, RKDS_2*) and Yan *et al.* (*YCM*) do certainly not match the experimentally obtained spectrum as the energies of the whiteline (~9008eV) and the peak position in the fine structure region at ~9045eV are significantly shifted to lower energies (dashed lines, Fig. 2). Moreover, the overall shape, i.e. the number and relative amplitudes of the peaks clearly differs from the experimental observation. This results from a local atomic structure around the absorbing Cu atoms, which is considerable different to the present structure and, thus, the *RKDS* and *YCM* models can be clearly rejected.

Regarding the simulated spectra of the *PW* and *M_1* (*M_2*) models, the near edge region (8980-9010eV) shows an almost identical curve shape, whereas significant deviations occur for the peak positions in the fine structure region (9045-9100eV). Compared to the experimentally obtained spectrum, a slight energy shift of the whiteline (~9008eV) to higher and lower energies can be identified for the *M_1* (*M_2*) and *PW* models, respectively. The *M_1* (*M_2*) models indicate a better agreement with the fine structure peak at ~9045eV as this peak position is slightly shifted to lower energies for the *PW* model. On the contrary, at energies between 9050eV and 9100eV the *PW* model rather reflects the overall shape and peak positions since an additional peak at ~9075eV can be identified for the *M_1* (*M_2*) models, which is clearly not present in the experimentally obtained spectrum. As for the whiteline (~9008eV) and the fine structure peak (~9045eV) in the *PW* model, the position of the peak minimum at ~9075eV is as well shifted to lower energies.

---

[1] The formation of either Cu-Mg co-clusters (or vacancy Cu-Mg complexes) [39] or GPB zones [1,2] at the very beginning of the precipitation sequence is controversially discussed. However, the evaluation of the structures occurring at the very early stages in Al-Cu-Mg alloys is not subject of the present work but will be part in forthcoming publications.

In general, slight deviations in the bonding distances to the nearest neighbours of the probed atoms cause a phase shift of the interfered spherical photoelectron wave, whereas variations in coordination numbers and/or types of atoms due to a different space group result in a different interference pattern. Hence, slight energy shifts of the fine structure peaks in the *PW* model mainly originate from differences in the lattice parameters and/or atomic positions, whereas this additional peak at ~9075eV of the *M_1* (*M_2*) models result from an incorrect arrangement of the atoms surrounding Cu. Based on these observations, we are not in favour with the structure proposed by Mondolfo.

The refined *PW* models proposed by Heying *et al.* (*HHP*) and Styles *et al.* (*SHCDB_3*, *SHCDB_4*) indicate an even better agreement, since the peak positions in the fine structure region fully match the experimental data (Fig. 2b). However, deviations occur now to higher energies in the whiteline region. On the contrary, the *SHCDB_1* (*SHCDB_2*) refinements show no significant changes to the original *PW* model.

As shown by Baur & Gerold, a considerable amount of isolated Cu remains in solid solution during the formation and growth of Guiner-Preston (GP) zones in Al-Cu alloys and strongly depends on the aging temperature and Cu concentration [41]. Hence, we assume that the experimental curve, especially in the near edge region (8980-9010eV) might be rather reflected by a superposition of solute Cu atoms with the refined structures based on the model proposed by Perlitz & Westgren (see also [35]). The experimental curve could as well be matched by a superposition or distribution of various lattice parameters and/or atomic positions, where those given by Heying *et al.* and Styles *et al.* represent potential mean values. Furthermore, coexisting metastable phases (GPB (II) zones), as possibly present in the investigated state (implied by the endothermal peak at ~275°C) might also affect the shape of the experimental XAFS spectrum in case that their volume fraction is above the detection limit.

Finally, our study is the first to support the widely accepted *PW* model as the only correct crystal structure of the S phase in pure Al-Cu-Mg alloys by XAFS measurements.

## 4 Conclusion

In this study, we used for the first time XAFS to verify the *true* S phase structure by comparing the experimental curve of an artificially aged Al-Cu-Mg alloy (200°C, 1d) to simulated spectra of the structures proposed by Perlitz & Westgren, Mondolfo, Yan *et al.* and Radmilovic & Kilaas. Only models which clearly refer to the intermediate (S') or equilibrium S phase were considered. To assure that mainly S phase precipitates are present in the studied sample,

DSC measurements were carried out before. Finally, we could confirm that the widely accepted *PW* model is the only correct structure of the S phase as the other models indicated significant deviations from the overall experimental curve shape. Comparisons to refined versions of the *PW* models showed that slight modifications in the lattice parameters could even lead to a better match in the peak positions. Thus, we have shown that XAFS provides a powerful tool for structural investigations on Al alloys.

**Acknowledgments**

This project was funded by the German Research Foundation (DFG) (STA-527/5-1). We acknowledge the European Synchrotron Radiation Facility (ESRF, Grenoble, France) for provision of synchrotron radiation facilities and we would like to thank Manuel Monte Caballero for assistance in using beamline ID24 of ESRF.

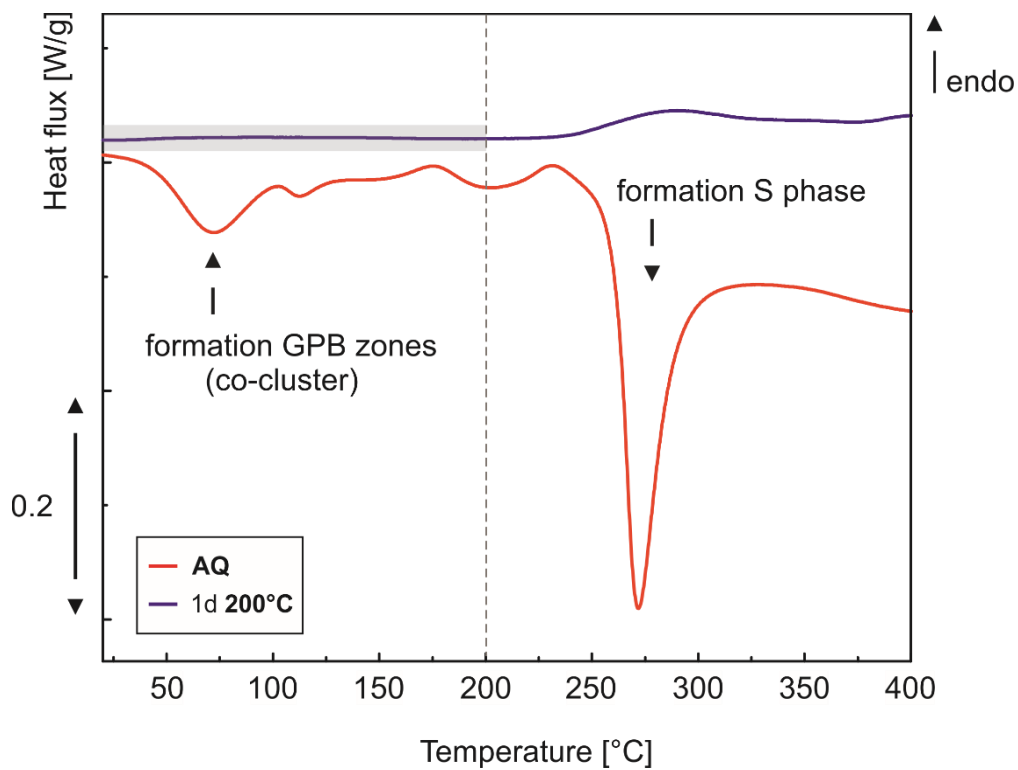

**Fig. 1** DSC thermograms of the Al – 2.2Cu – 1.6Mg (wt.%) alloy acquired for the as-quenched (AQ, red curve) and artificially aged sample state (200°C for 1d, blue curve). The S formation peak (250-300°C) has completely disappeared in the studied sample state, which indicates the transformation of the majority of metastable phases (GPB-II/S'', S') into the equilibrium S phase.

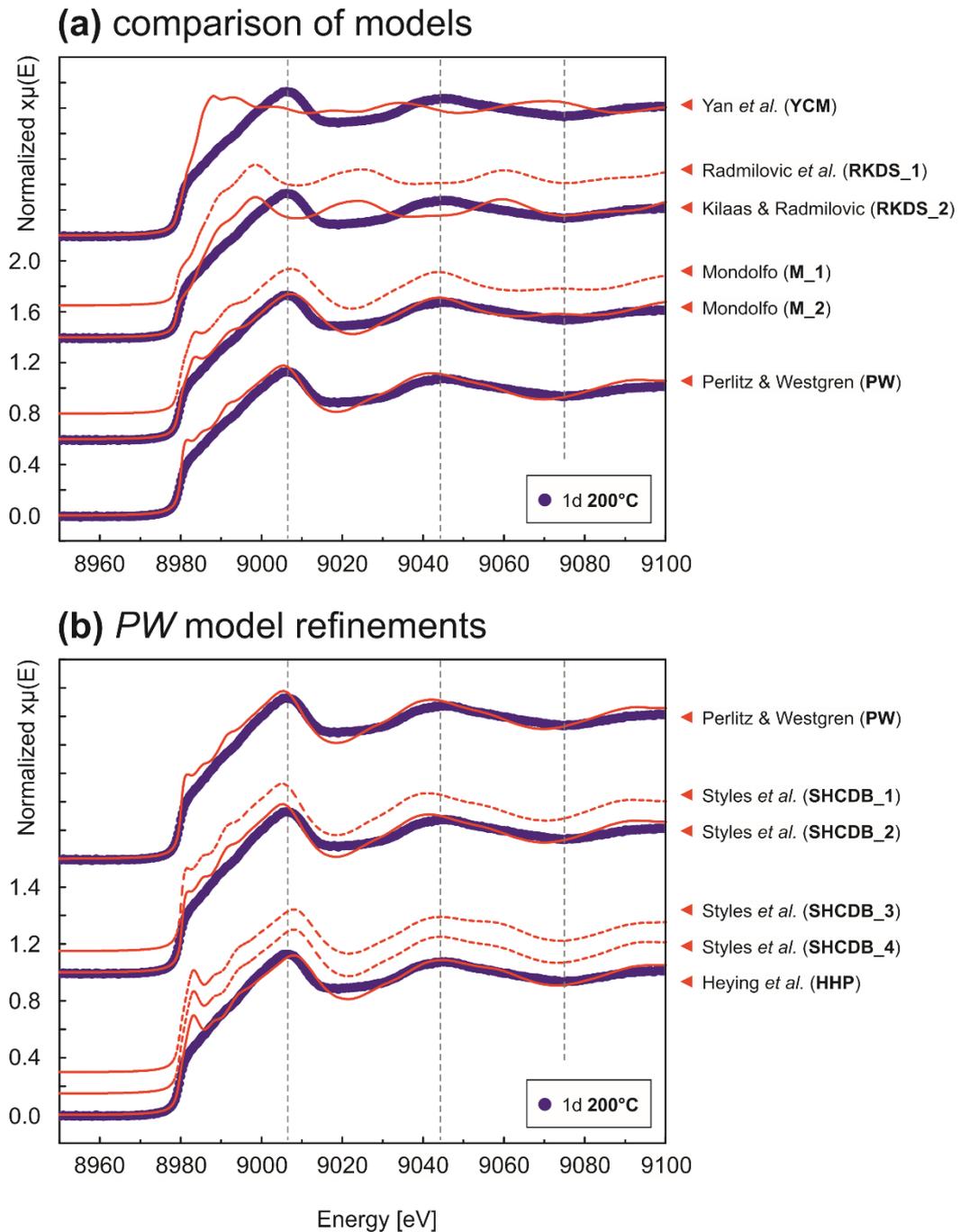

**Fig. 2** Comparison between the experimentally obtained spectrum (200°C for 1d, blue dots) and the simulations by FDMNES (red curves), which refer to **(a)** the proposed S phase models by Perlitz & Westgren (PW) [14], Mondolfo [15], Radmilovic & Kilaas *et al.* [4,17] and Yan *et al.* [16] and to **(b)** the structure refinements of the *PW* model proposed by Styles *et al.* [10,11] and Heying *et al.* [12].

| notation | exp. methods of observation | reference (date of observation) |
| --- | --- | --- |
| **PW** | single crystal XRD | Perlitz, Westgren (1943) [14] |
| **M_1** | single crystal XRD | Mondolfo (1976) [15] |
| **M_2** | single crystal XRD | Mondolfo (1976) [15] |
| **YCM** | TEM | Yan *et al.* (1990) [16] |
| **RKDS_1** | HRTEM | Radmilovic *et al.* (1999) [4] |
| **RKDS_2** | HRTEM | Kilaas & Radmilovic (2001) [17] |
| **HHP** | single crystal + powder XRD | Heying *et al.* (2005) [12] |
| **SHCDB_1** | synchrotron X-ray powder diffraction + TEM | Styles *et al.* (2012) [10] |
| **SHCDB_2** | synchrotron X-ray powder diffraction + TEM | Styles *et al.* (2012) [10] |
| **SHCDB_3** | synchrotron X-ray powder & neutron diffraction | Styles *et al.* (2015) [11] |
| **SHCDB_4** | synchrotron X-ray powder & neutron diffraction | Styles *et al.* (2015) [11] |

**Table 1** Chronologically sorted by their date of observation (from top to bottom): All proposed crystal structures as can be found in the literature for the intermediate (S') and equilibrium S phase. The second column assigns the experimental methods on which the proposal is based on.

| notation | unit cell (UC) (space group) | lattice parameters (a, b, c) [Å] | x | y | z | occupancy |
|---|---|---|---|---|---|---|
| **PW** | orthorhombic (Cmcm) | (4.0, 9.23, 7.14) | 0 | 0.778 | 0.25 | 100% Cu |
| | | | 0 | 0.072 | 0.25 | 100% Mg |
| | | | 0 | 0.356 | 0.056 | 100% Al |
| **M_1** | triclinic (P1) | (4.0, 9.25, 7.18) | 0 | 0.928 | 0.25 | 100% Cu |
| | | | 0.5 | 0.428 | 0.25 | 100% Cu |
| | | | 0 | 0.222 | 0.75 | 100% Cu |
| | | | 0.5 | 0.722 | 0.75 | 100% Cu |
| | | | 0 | 0.356 | 0.056 | 100% Al |
| | | | 0.5 | 0.856 | 0.056 | 100% Al |
| | | | 0 | 0.644 | 0.556 | 100% Al |
| | | | 0.5 | 0.144 | 0.556 | 100% Al |
| | | | 0 | 0.356 | 0.444 | 100% Al |
| | | | 0.5 | 0.856 | 0.444 | 100% Al |
| | | | 0 | 0.644 | 0.944 | 100% Al |
| | | | 0.5 | 0.144 | 0.944 | 100% Al |
| | | | 0 | 0.222 | 0.25 | 100% Mg |
| | | | 0.5 | 0.722 | 0.25 | 100% Mg |
| | | | 0 | 0.928 | 0.75 | 100% Mg |
| | | | 0.5 | 0.428 | 0..75 | 100% Mg |
| **M_2** | triclinic (P1) | (4.0, 9.25, 7.14) | see *M_1* structure | | | |
| **YCM** | orthorhombic (Pmm2) | (4.0, 4.61, 7.18) | 0 | 0.5 | 0.75 | 100% Cu |
| | | | 0 | 0 | 0 | 100% Al |
| | | | 0.5 | 0 | 0 | 100% Al |
| | | | 0 | 0.5 | 0.25 | 100% Mg |
| **RKDS_1** | orthorhombic (Cmcm) | (4.03, 9.3, 7.08) | 0 | 0.778 | 0.25 | 100% Mg |
| | | | 0 | 0.072 | 0.25 | 100% Cu |
| | | | 0 | 0.356 | 0.056 | 100% Al |

| Name | Unit cell (space group) | Lattice parameters (Å) | x | y | z | Occupancy |
|---|---|---|---|---|---|---|
| **RKDS_2** | orthorhombic | (4.03, 9.3, 7.08) | 0 | 0.765 | 0.25 | 100% Mg |
| | (Cmcm) | | 0 | 0.074 | 0.25 | 100% Cu |
| | | | 0 | 0.362 | 0.056 | 100% Al |
| **HHP** | orthorhombic | (4.0119, 9.265, 7.124) | 0 | 0.3558 | 0.0556 | 100% Al |
| | (Cmcm) | | 0 | 0.7801 | 0.25 | 100% Cu |
| | | | 0 | 0.0651 | 0.25 | 100% Mg |
| **SHCDB_1** | orthorhombic | (4.049, 9.25, 7.09) | | see *PW* structure | | |
| | (Cmcm) | | | | | |
| **SHCDB_2** | orthorhombic | (4.016, 9.27, 7.117) | | see *PW* structure | | |
| | (Cmcm) | | | | | |
| **SHCDB_3** | orthorhombic | (4.0478, 9.2552, 7.0948) | 0 | 0.3537 | 0.0538 | 100% Al |
| | (Cmcm) | | 0 | 0.7809 | 0.25 | 96% Cu |
| | | | 0 | 0.0624 | 0.25 | 91% Mg |
| **SHCDB_4** | orthorhombic | (4.0144, 9.2677, 7.1234) | 0 | 0.3553 | 0.0531 | 100% Al |
| | (Cmcm) | | 0 | 0.7808 | 0.25 | 100% Cu |
| | | | 0 | 0.067 | 0.25 | 99% Mg |

**Table 2** Space group, type of unit cell, lattice parameters, atomic sites and site occupancies of the proposed crystal structures for the intermediate (S') and equilibrium S phases as given in Table 1. These coordinates served as input for the XAFS spectra simulations with *FDMNES*.